\documentclass[nohyperref]{article}

\usepackage{microtype}
\usepackage{graphicx}
\usepackage{subfigure}
\usepackage{booktabs} 

\usepackage{hyperref}

\usepackage[accepted]{icml2022}

\usepackage{amsmath}
\usepackage{amssymb}
\usepackage{mathtools}
\usepackage{amsthm}

\usepackage[capitalize,noabbrev]{cleveref}

\theoremstyle{plain}

\theoremstyle{definition}

\theoremstyle{remark}

\usepackage[textsize=tiny]{todonotes}

\icmltitlerunning{BNNs for Low-Surface-Brightness-Galaxy Parameter Estimation}

\begin{document}

\twocolumn[
\icmltitle{Inferring Structural Parameters of Low-Surface-Brightness-Galaxies with Uncertainty Quantification using Bayesian Neural Networks}



\icmlsetsymbol{equal}{*}

\begin{icmlauthorlist}
\icmlauthor{Dimitrios Tanoglidis}{UChic,Kavli}
\icmlauthor{Aleksandra Ćiprijanović}{Fermi}
\icmlauthor{Alex Drlica-Wagner}{Fermi,UChic,Kavli}
\end{icmlauthorlist}

\icmlaffiliation{UChic}{Department of Astronomy \& Astrophysics, University of Chicago, Chicago IL 60637, USA}
\icmlaffiliation{Kavli}{Kavli Institute for Cosmological Physics, Chicago IL 60637, USA}
\icmlaffiliation{Fermi}{Fermi National Accelerator Laboratory, Batavia,
IL 60510, USA}


\icmlcorrespondingauthor{Dimitrios Tanoglidis}{dtanoglidis@uchicago.edu}

\icmlkeywords{Machine Learning, ICML}

\vskip 0.3in
]



\printAffiliationsAndNotice{}  

\begin{abstract}
Measuring the structural parameters (size, total brightness, light concentration, etc.) of galaxies is a significant first step towards a quantitative description  of different galaxy populations. In this work, we demonstrate that a Bayesian Neural Network (BNN) can be used for the inference, with uncertainty quantification, of such morphological parameters from simulated  low-surface-brightness galaxy images. 
Compared to traditional profile-fitting methods, we show that the uncertainties obtained using BNNs are comparable in magnitude, well-calibrated, and the point estimates of the parameters are closer to the true values. Our method is also significantly faster, which is very important with the advent of the era of large galaxy surveys and big data in astrophysics.
\end{abstract}

\section{Introduction}
\label{Intro}

Despite their morphological diversity and complexity, the approximate light distribution of galaxies can be well-described by analytic fitting functions with a limited number of free parameters, such as their orientation, size (radius), light concentration, total brightness, etc. Measurements of these parameters allow for a quantitative comparison of different galaxy populations and the derivation of empirical scaling relations \citep[e.g.,][]{Scaling_2007}, which in turns facilitates the testing of galaxy formation models. 

Traditionally, these parameters are measured using galaxy profile fitting software (two widely used options being \texttt{GALFIT}; \citet{Peng_2002} and \texttt{Imfit}; \citet{Erwin_2015}) that performs a $\chi^2$ minimization between the chosen analytic model and a given galaxy image, to derive the best-fit parameters. Despite their success, these codes have their limitations, too: they are not optimized to fit a large number of galaxies quickly, and they usually require some manual intervention (for example, the selection of good initial model parameters). The low speed is an even more  significant problem if one wants to obtain accurate estimates of the uncertainties associated with those measurements, for example via bootstrap resampling, or by using a Markov-Chain Monte Carlo (MCMC) approach to sample the parameter posterior distribution. 

Large galaxy surveys, such as the Dark Energy Survey (DES)\footnote{\url{https://www.darkenergysurvey.org/}} and the upcoming Legacy Survey of Space and Time (LSST)\footnote{\url{https://https://www.lsst.org/}} on the Vera C.\ Rubin Observatory, observe hundreds of millions (the former) to tens of billions (the latter) of galaxies. With the advent of these surveys, fast, automated, and reliable methods for measuring the structural parameters of galaxies are needed. Deep learning methods are well-suited to tackle this problem since, once trained, they are able to make predictions on new, unseen, examples very quickly.

Indeed, several works \citep[e.g.,][]{Tucillo_2018,Aragon_Calvo_2020,Li_2022} have demonstrated that Convolutional Neural Networks (CNNs), trained on simulated galaxy images are significantly faster, and almost as accurate as the standard profile fitting methods in predicting galaxy parameters. However, those works used standard, deterministic, neural networks, that output single-point estimates and thus are unable to quantify the uncertainty associated with their predictions.

A rigorous uncertainty quantification is imperative for studies of challenging, low signal-to-noise objects, such as low-surface-brightness galaxies (LSBGs). LSBGs, defined as galaxies with a central brightness at least a magnitude fainter than that of the ambient dark sky, are challenging to observe and galaxy surveys have only recently started to produce large LSBG catalogs \citep{Greco_2018,Tanoglidis_2021,Zaritsky_2022}, althought they are expected to dominate the galaxy population. LSBGs are a target of future surveys, in the quest of understanding the galaxy formation process in a relatively unexplored regime. 

In this work, we explore the use of Bayesien Neural Networks \citep[BNNs; e.g.,][]{BNNs_2020} for the problem of LSBG structural parameter estimation with simultaneous uncertainty quantification. BNNs output posterior probability distributions instead of point estimates for their predictions, and thus they naturally offer a way to quantify the uncertainties associated with neural network predictions. Specifically, we use a simulated dataset of DES-like LSBGs, to train, validate, and test a convolutional BNN model and compare the speed and accuracy of its predictions with those obtained using \texttt{pyImfit}\footnote{\texttt{pyImfit} is a Python wrapper around \texttt{Imfit}.}, for a single-component Sérsic light-profile model.

Our code and simulated data related to this work are available at: \url{https://github.com/dtanoglidis/BNN_LSBGs_ICML}.

\section{Simulated Data}
\label{sec: Data}
 
\begin{figure}[ht]
\vskip 0.1in
\begin{center}
\centerline{\includegraphics[width=0.91\columnwidth]{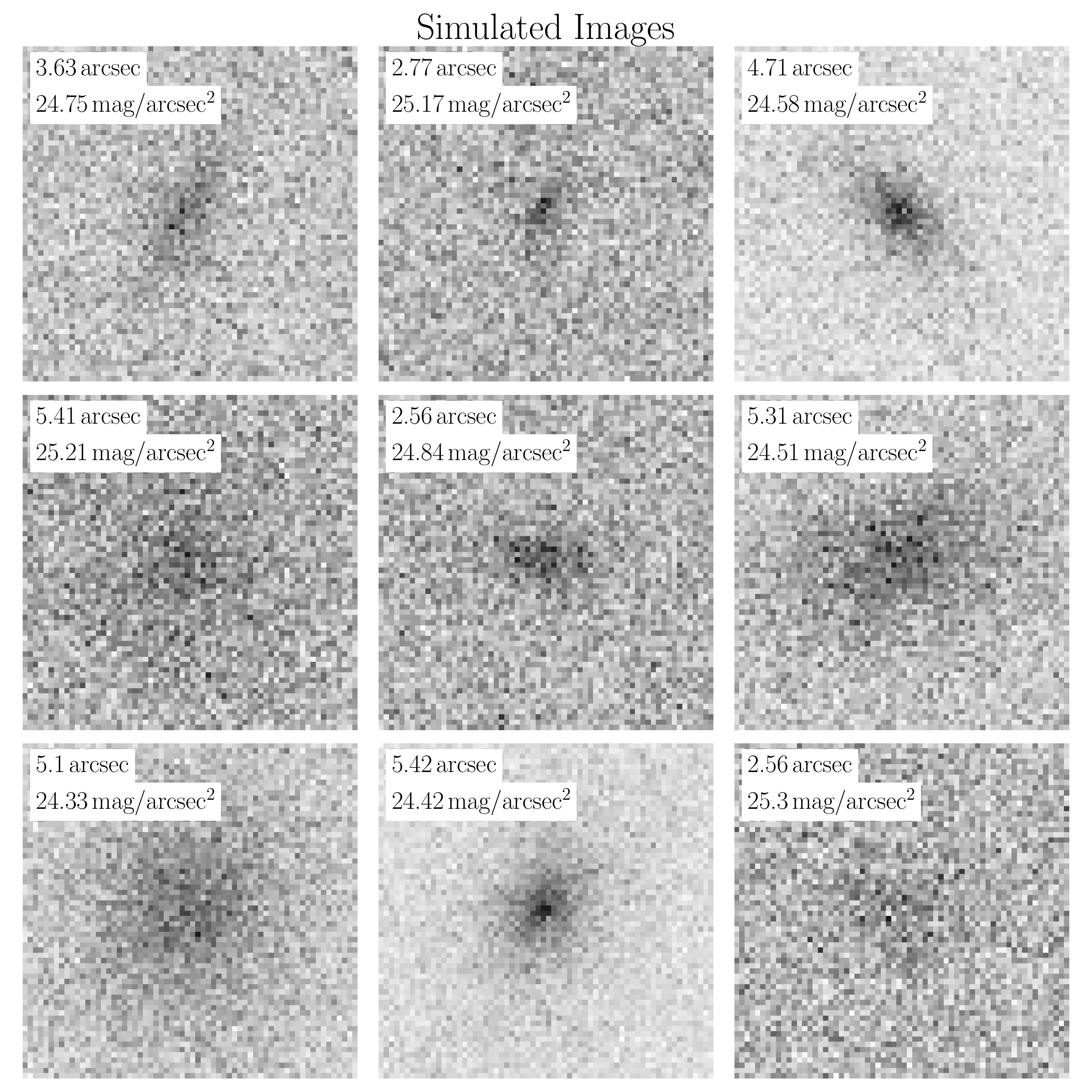}}
\caption{Examples of simulated LSBG images. The inset text refers to the effective radius, $r_e$, (top) and the surface brightness, $I_e$ (bottom).}
\label{Example_Dataset}
\end{center}
\vskip -0.2in
\end{figure}
We use \texttt{PyImfit} to create a simulated dataset of 170,000 LSBG images. Each image has dimensions $64\times64$ pixels, and it has two components: a uniform background $I(x,y)=I_{\mbox{\scriptsize{sky}}}$, and a Sérsic function \citep{Sersic_1963} that describes the surface-brightness profile of the galaxies: 

\begin{equation}
I_{\mbox{\scriptsize{gal}}}(r) = I_e \exp \left\{ -b_n\left[\left(\frac{r}{r_e}\right)^{1/n}-1\right] \right\},
\end{equation}
where $r=(x^2+y^2/q^2)^{1/2}$, $(x,y)$ are the coordinates with origin at the center of the image, and $q$ is the axis ratio. The axis ratio is connected to the ellipticity as $q=1-\epsilon$. The other free parameters of the model are: the effective half-light radius, $r_e$, the surface brightness at the effective radius, $I_e$, the Sérsic index, $n$, that controls the shape of the light distribution, and the position angle that defines the orientation of the galaxy profile. As for the value of $b_n$ (not a free parameter of the model), \texttt{pyImfit} uses the approximation by \citet{Ciotti_1999}.

We want our simulated images to resemble real LSBG images. For that reason we sample parameters uniformly, from a range that roughly corresponds to the bulk of the LSBGs discovered by DES, as described in \citet{Tanoglidis_2021}: \begin{itemize}
\setlength\itemsep{-0.45em}
    \item Position angle, PA $\in [0,180]$ degrees,
    \item Ellipticity, $\epsilon \in [0.05,0.7]$,
    \item Sérsic index, $n \in [0.5,1.5]$,
    \item Surface brightness, $I_e \in [24.3,25.5]$ mag/arcsec$^2$,
    \item Effective radius, $r_e \in [2.5,6.0]$ arcsec.
\end{itemize}
Furthermore, we assume a pixel to angular scale conversion 1 pix = 0.263 arcsec (as in DES), and background sky surface brightness $I_{\mbox{\scriptsize{sky}}} = 22.23$ mag/arcsec$^2$ \citep{Neilsen_2016}.
At each pixel we randomly assign a number of photons (counts) drawn from a Poisson distribution with mean the one predicted from the total surface brightness model $I_{\mbox{\scriptsize{tot}}}=I_{\mbox{\scriptsize{gal}}}+I_{\mbox{\scriptsize{sky}}}$.
In Fig. \ref{Example_Dataset} we present a small subset of the simulated galaxy images. 

\section{Model}

\subsection{Bayesian Neural Networks (BNNs)}

Standard neural networks, once trained, output a single point estimate prediction every time the same example is presented to the network. Thus, \textit{deterministic} neural networks are unable to capture uncertainties in their predictions. In BNNs, the single weights are being replaced by appropriate probability distributions, that can be subsequently used to provide a measure of how (un)certain a model is in its predictions. 

Given a training dataset, ${\cal{D}}=(\mathbf{x},\mathbf{y})$, training a BNN consists of finding the posterior distribution of the weights, $p(w|{\cal{D}})$. The exact evaluation of this posterior distribution is computationally intractable, since it requires multi-dimensional integration over the weight values. One approach (variational inference; VI) is to approximate the true posterior with a distribution $q(w|\theta)$ of a known form (e.g., a Gaussian), with free parameters $\theta$ to be learned. The goal is to select the parameters to minimize the difference between the true and approximate posteriors. It can be shown that this is equivalent to minimizing the (negative) evidence lower bound (ELBO) loss function \citep[e.g.,][]{Bayes_by_Backprop}:
\begin{equation}
{\cal{L}}({\cal{D}},\theta) =  \mathbb{E}_{q(w|\theta)}\left[\log q(w|\theta) - \log p(w)p({\cal{D}}|w)\right].
\end{equation}
Once trained, random weights can be drawn from the approximate posterior $q(w|\theta)$, and the network can give predictions on new examples presented to it.

\subsection{Implementation and Training}

Our BNN architecture consists of five (probabilistic) 2D convolutional layers, each one followed by a Max Pooling layer, and a (probabilistic) dense layer following the convolutional part. The model output is a multivariate ($n=5$ dimensions, equal to the five free parameters of the Serśic model described in Sec. \ref{sec: Data}) Gaussian, with full covariance. For the interested reader, we present a schematic overview and more details about the architecture in Appendix \ref{Sec: Architecture}.

We implement our network using the \texttt{Keras} and \texttt{Tensorflow Probability (TFP)} Python libraries. Before training, we randomly split the full simulated dataset to a training (150k), a validation (10k), and a test (10k) set. We perform training with a learning rate $\eta =0.2$ (\texttt{Adadelta} optimizer), for 150 epochs, using a batch size of 64. During training we observed no signs of overfitting. We utilized the 25 GB high-RAM Nvidia P100 GPUs available through the Google Colab Pro. The training took approximately three hours to complete.

\section{Results}

\subsection{Parameter posteriors}

\begin{figure}[!ht]
\vskip 0.2in
\begin{center}
\subfigure[]{\includegraphics[width=0.86\columnwidth]{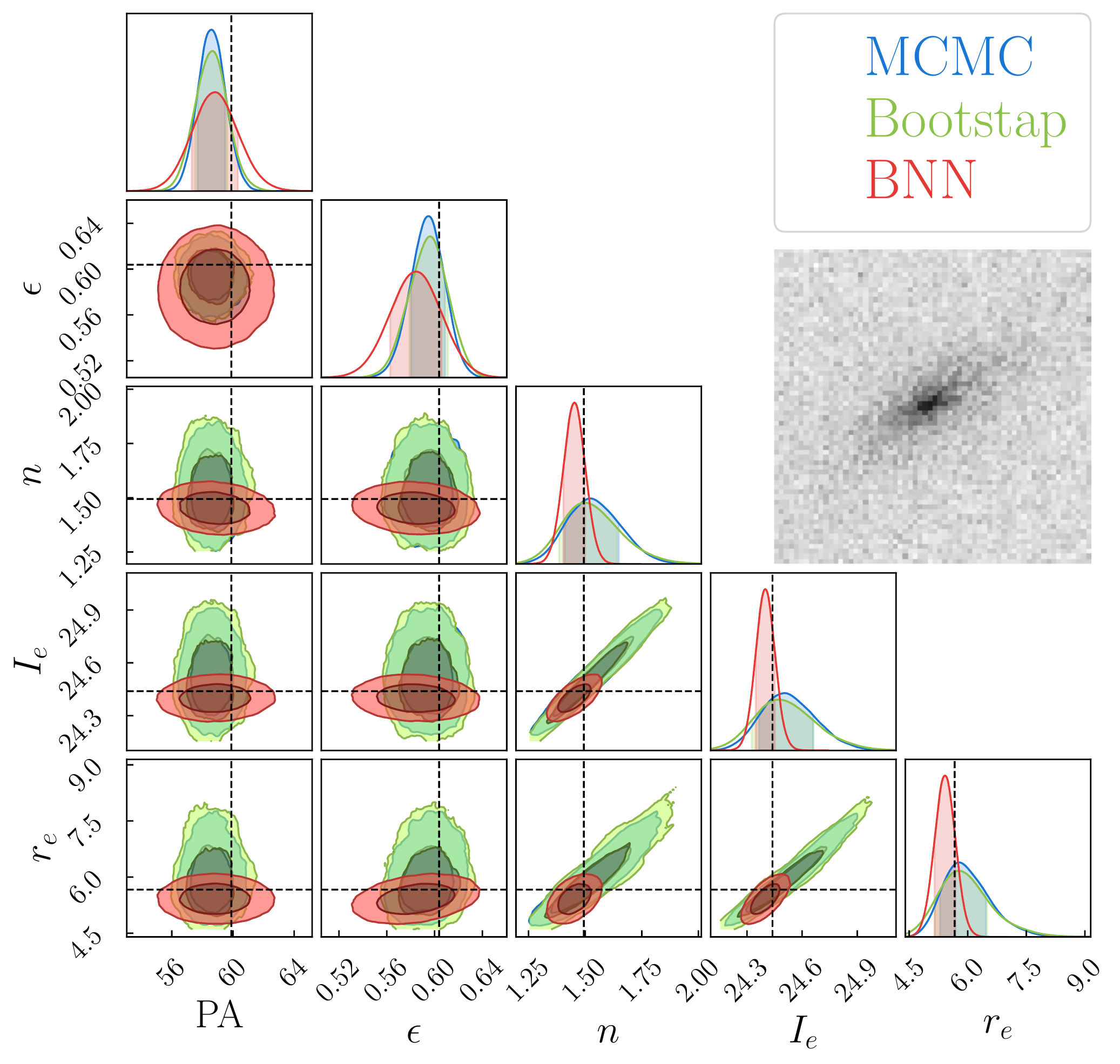}}
\subfigure[]{\includegraphics[width=0.86\columnwidth]{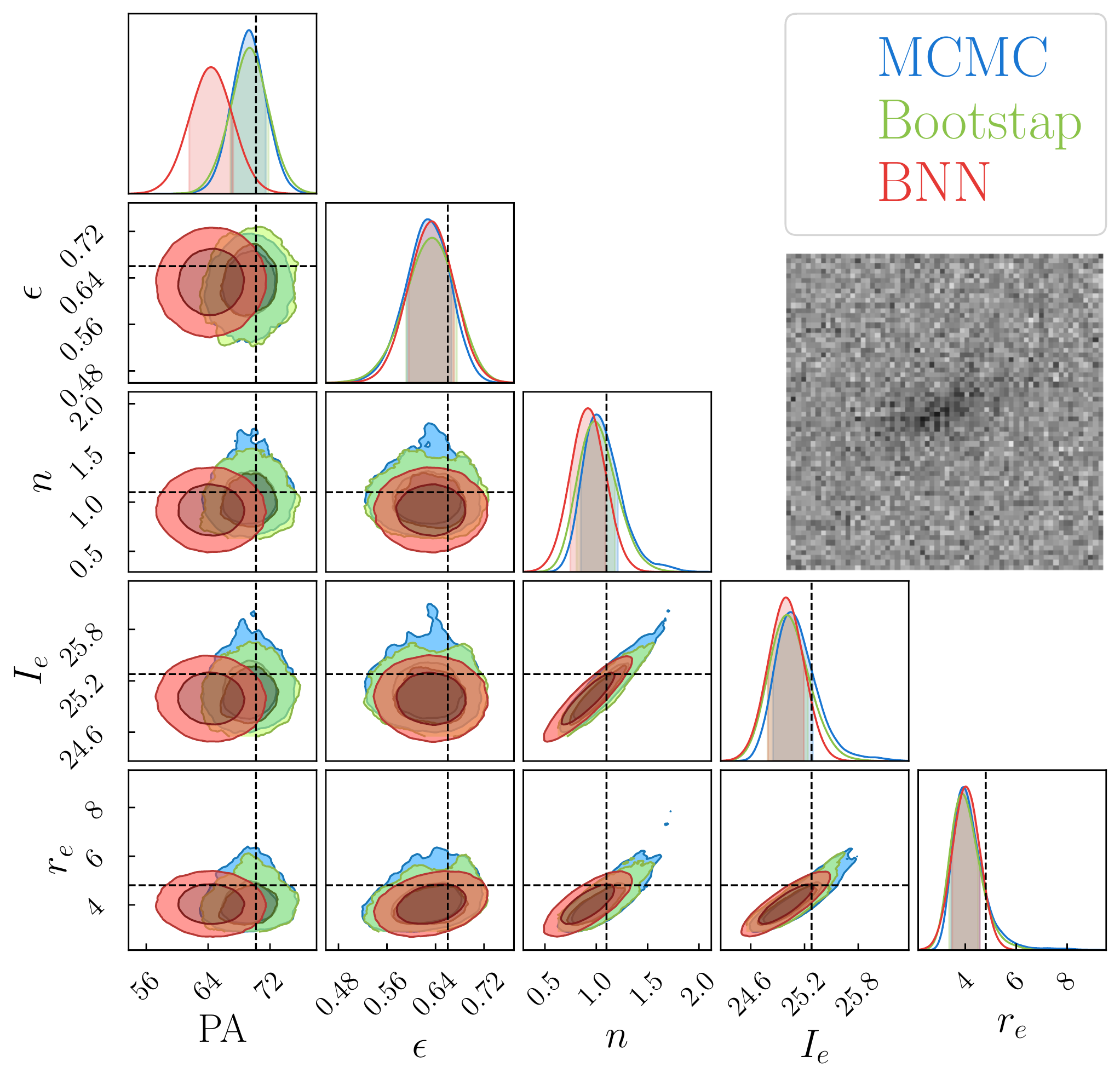}}
\caption{Predicted posterior distributions of structural parameters for an LSBG in the bright end ($I_e = 24.4$ mag/arcsec$^2$, panel (a)), and one in the faint end ($I_e = 25.3$ mag/arcsec$^2$, panel (b)) of the surface brightness range we consider in this work. The dashed lines indicate the true parameter values. The plots were created using \texttt{ChainConsumer} \citep{Hinton2016}.}
\label{Example_Plots}
\end{center}
\vskip -0.25in
\end{figure}

In Fig. \ref{Example_Plots} we present the predicted posterior distributions for the five parameters of the Sérsic model, described in Sec. \ref{sec: Data}, for a simulated galaxy at the bright end of the surface brightness distribution ($I_e = 24.4$ mag/arcsec$^2$, panel (a)), and one at the faint end ($I_e = 25.3$ mag/arcsec$^2$, panel (b)). We present the predictions of the BNN model (red contours) and those from the \texttt{pyImfit} model, using two different estimation methods: bootstrap resampling (green contours), and Markov chain Monte Carlo (MCMC; blue contours).

To get the BNN posteriors, we stack together the output distributions from 400 forward passes (predictions) of the model, for each one of the LSBG images. We see that the constraints on the parameters are tighter (as in the case of the brighter LSBG) or comparable (as in the case of the fainter LSBG) to those obtained using \texttt{pyImfit}. Although we present only two examples here, we have confirmed that this is true for a larger number of randomly selected LSBG images.

In terms of speed, obtaining the posterior distribution for each one of the examples using BNNs was significantly faster than running MCMC ($\sim 1$ minute vs $\sim$ 6 minutes), and comparable in time to the bootstrap method. The real gain in time  comes when one wants to process a large number of galaxy images simultaneously; for example, obtaining full posterior estimates for 1000 LSBGs using BNNs takes $\sim 7$ minutes on our machine, while processing the same number of images using \texttt{pyImfit} and bootstrap resampling would had taken $\sim 16$ hours.

\subsection{Calibration}
\label{sec: calibration}

\begin{figure}[ht]
\vskip 0.1in
\begin{center}
\centerline{\includegraphics[width=0.83\columnwidth]{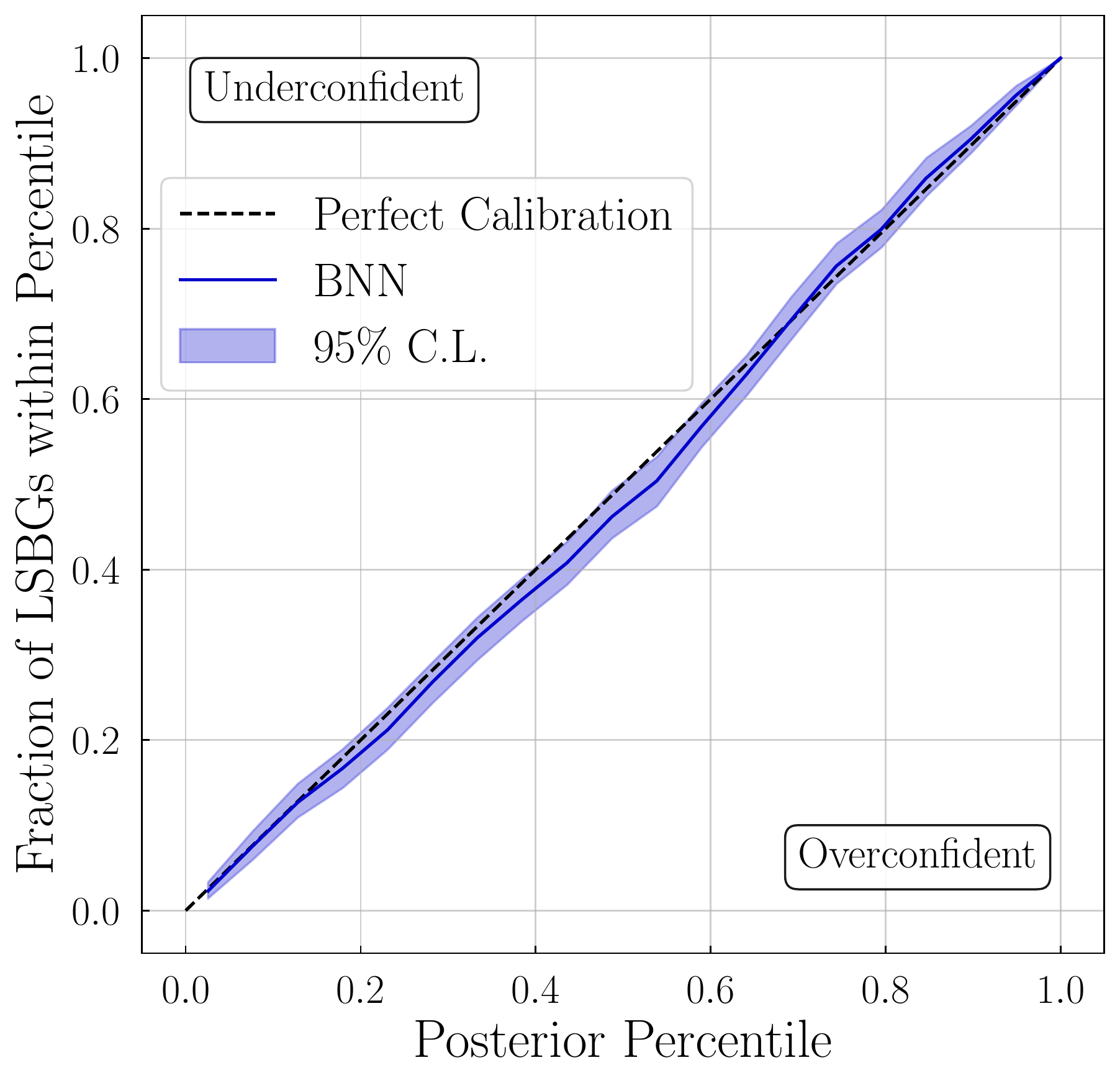}}
\caption{Calibration curve of the BNN posterior for the effective radius parameter, $r_e$. Within the statistical uncertainty, the BNN curve agrees with the diagonal line that indicates perfect calibration.}
\label{Calibration_Plot}
\end{center}
\vskip -0.2in
\end{figure}

We have showed that BNNs fit Sérsic model parameters with tighter or similar uncertainties to those produced by \texttt{pyImfit}. In order to be interpreted as confidence intervals, we have to demonstrate that a $x\%$ interval contains the true value $x\%$ of the time -- in other words, that the posterior is well-calibrated.

To investigate that, we consider the parameter posterior predictions on 1000 simulated LSBGs drawn from the test set. Following \citet{Wagner_Carena_2021,Park_2021}, we calculate, at different posterior percentile levels, the fraction of LSBGs with true values within the limits of that percentile interval. For well-calibrated posteriors, those two quantities (percentile and fraction) should be equal. Here, we show only the (marginalized) posterior for the effective radius, $r_e$. However, similar results are found for the other parameters; the interested reader can see those plots in Appendix \ref{Sec: Cal_Plots}. As we can see in Fig. \ref{Calibration_Plot}, the posterior produced by the BNN is statistically consistent with being perfectly calibrated (the error band was calculated by performing a bootstrap resampling of the test set used for the calculations).

\subsection{Comparison of point estimates}
\label{sec: point_estimates}

\begin{figure}[ht]
\vskip 0.1in
\begin{center}
\centerline{\includegraphics[width=0.83\columnwidth]{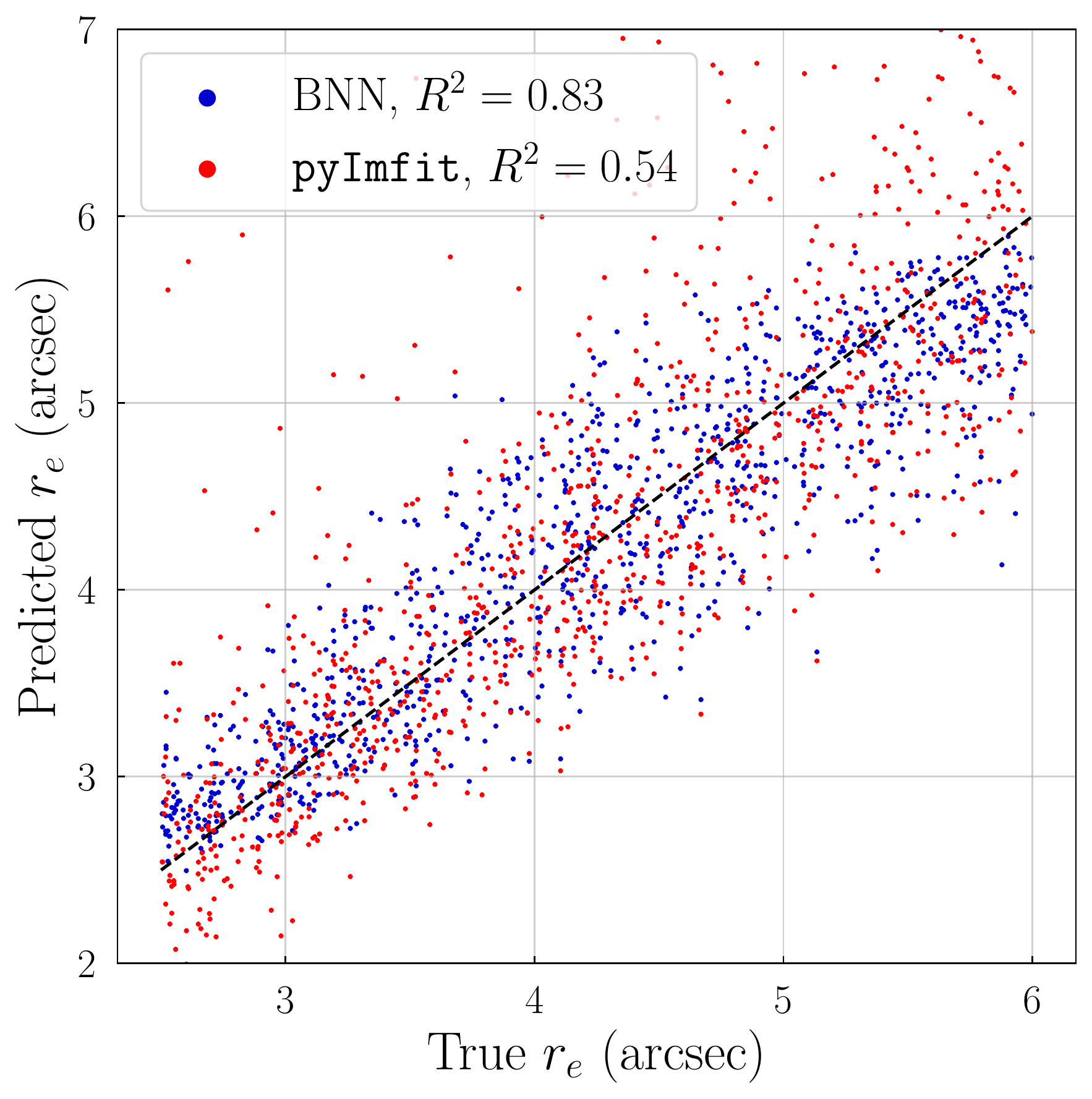}}
\caption{Comparison of the effective radius parameter, $r_e$, predictions from the BNN model (blue dots) and from \texttt{pyImfit} (red dots) versus the true values.}
\label{Comparison}
\end{center}
\vskip -0.2in
\end{figure}

We have demonstrated that the BNN model outputs well-calibrated errors, and we have seen examples where we compared the output parameter posteriors from the BNN and the \texttt{pyImfit} algorithm. We now compare the point estimate (mean) prediction from the BNN with the best fit parameter output from \texttt{pyImfit}. 
 
In Fig. \ref{Comparison} we plot the true value of the effective radius parameter (for the same 1000 simulated LSBGs as in the previous section) vs the predicted one, using both methods. The point estimates produced by the BNN method tend to be closer to the true value, as indicated by the higher coefficient of determination ($R^2 = 0.83$ for BNN vs $R^2 = 0.54$ for \texttt{pyImfit}), and it performs significantly better for higher effective radius values. In Appendix \ref{Sec: Cal_Plots} we present similar plots for the other Sérsic model parameters.

\section{Discussion and Future Work}

We have used a Bayesian Neural network model to predict structural parameters of LSBGs in simulated galaxy images. We compare the posterior parameter predictions from the BNN method with those from a profile-fitting algorithm (\texttt{pyImfit}) for simulated LSBGs and we show that the BNN gives comparable or even tighter parameter constraints.

We furthermore show that the uncertainties estimated using the BNN method are well calibrated, and that, for a sample of simulated LSBGs, the BNN gives better point-estimate parameter predictions (higher coefficient of determinations) compared to those from \texttt{pyImfit}.

A significant strength of our BNN method is its speed. For example, it can predict the full posterior distribution of the five-parameter Sérsic model for 1000 LSBGs images within $\sim 7$ minutes (on the machine used here); using \texttt{pyImfit} and the bootstrap resampling methods to get parameter constraints for the same number of images, would require $\sim 16$ hours.  

An important next step, which we plan to address in future work, is to test the performance of our method on real LSBG images and investigate ways to improve it if necessary (for example by re-training on real data, as in \citet{Tucillo_2018}).
Indeed, the case presented here, where both training and testing was done on very simple simulated data, can be significantly different from real data. However, some preliminary investigation on applying the model trained here on real LSBGs has shown promising results.

Other areas of future investigation include testing different BNN architectures, testing the performance of the model on data outside of the training range, and a more rigorous comparison of the performance (parameter constraints and speed) between BNNs and \texttt{pyImfit}.

\section*{Acknowledgements}

We would like to thank Alex Ji, Brian Nord, Ji Won Park, and the anonymous referees for helpful suggestions. 

We acknowledge the Deep Skies Lab as a community of multi-domain experts and collaborators
who’ve facilitated an environment of open discussion, idea-generation, and collaboration. This
community was important for the development of this project.

This material is based upon work supported by the National Science Foundation under Grant No.\
AST-2006340. This work was partially funded by Fermilab LDRD 2018-052.

A. Ćiprijanović is partially supported by the High Velocity Artificial Intelligence grant as part of the Department of
Energy High Energy Physics Computational HEP
sessions program.

This work was supported by the University of Chicago and the Department of Energy
under section H.44 of Department of Energy Contract No. DE-AC02-07CH11359 awarded to Fermi
Research Alliance, LLC.

\nocite{langley00}

\bibliography{example_paper}
\bibliographystyle{icml2022}

\newpage
\appendix
\section{BNN Architecture}
\label{Sec: Architecture}

\begin{figure}[ht]
\vskip 0.2in
\begin{center}
\centerline{\includegraphics[width=0.85\columnwidth]{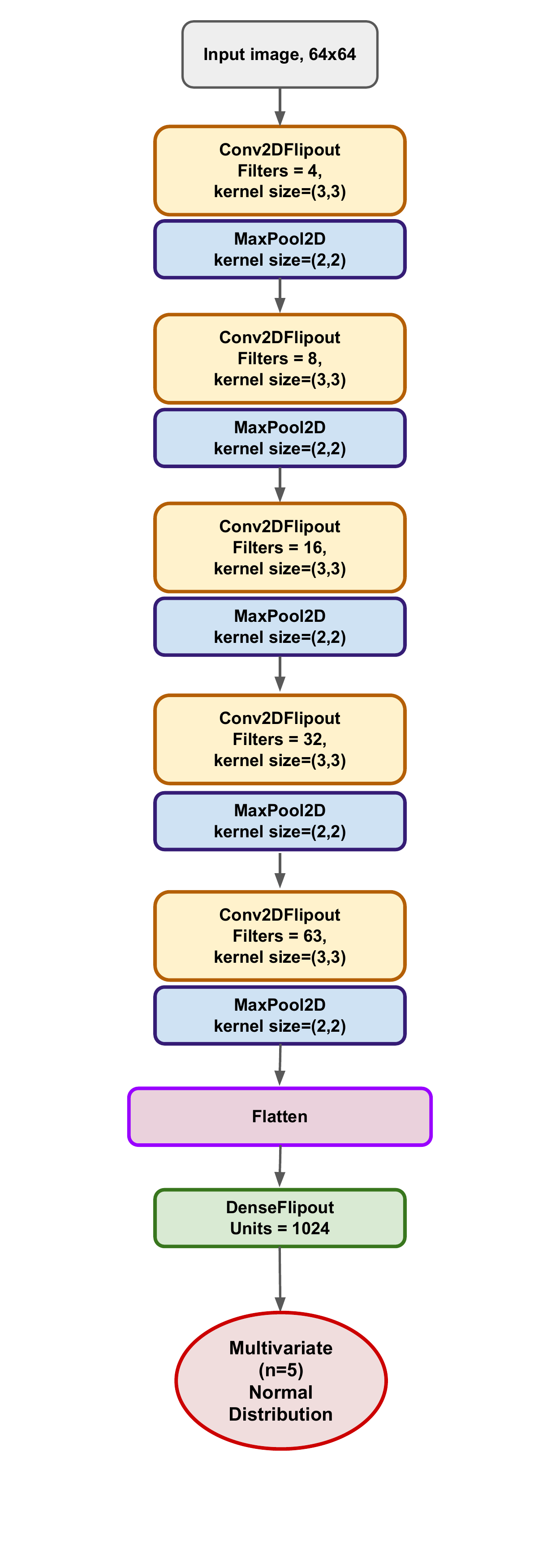}}
\caption{Schematic representation of the BNN architecture used in this work.}
\label{BNN_architecture}
\end{center}
\vskip -0.2in
\end{figure}

In Fig. \ref{BNN_architecture} we present a schematic overview of the BNN architecture used in this work. As we mentioned in the main text, we used the \texttt{Keras}\footnote{\url{https://keras.io/}} library on a \texttt{TensorFlow}\footnote{\url{https://www.tensorflow.org/}} backend, and the \texttt{Tensorflow Probability}\footnote{\url{https://www.tensorflow.org/probability}} extension of it, for the probabilistic layers.

The architecture consists of five probabilistic convolutional layers (\texttt{convolution2DFlipout}, provided by \texttt{TensorFlow Probability}). The number of filters and the kernel size used in each layer can be seen in the figure. 
Each convolutional layer is followed by a Max Pooling layer. After flattening we have a dense layer (\texttt{DelseFlipout}). The output is a multidimensional Gaussian (the five parameters our model tries to learn), with full covariance that allows to capture the correlations between the parameters. 

\onecolumn
\section{Calibration and Point Estimate Comparison Plots}
\label{Sec: Cal_Plots}

In Sec. \ref{sec: calibration}, we presented the calibration plot for the BNN posterior for the effective radius, $r_e$, and the comparison of the BNN point-estimate predictions with those coming from \texttt{pyImfit}, for the same parameter. We showed that the BNN errors are well-calibrated, and that the BNN point-estimates are closer to the true value compared to those of the \texttt{pyImfit} for the effective radius parameter. 

The figures presented in this Appendix demonstrate that these conclusions hold for other model parameters, too, with the notable exception of the position angle (PA). For the PA,  the network seems to be confused by the artificial discontinuity at the 0 to 180 degrees boundary (see Fig. 7d). In future iterations of this work, we plan to remove this discontinuity by reparameterizing the position angle.

\begin{figure}[ht]
\vskip 0.1in
\begin{center}
\subfigure[]{\includegraphics[width=0.37\textwidth]{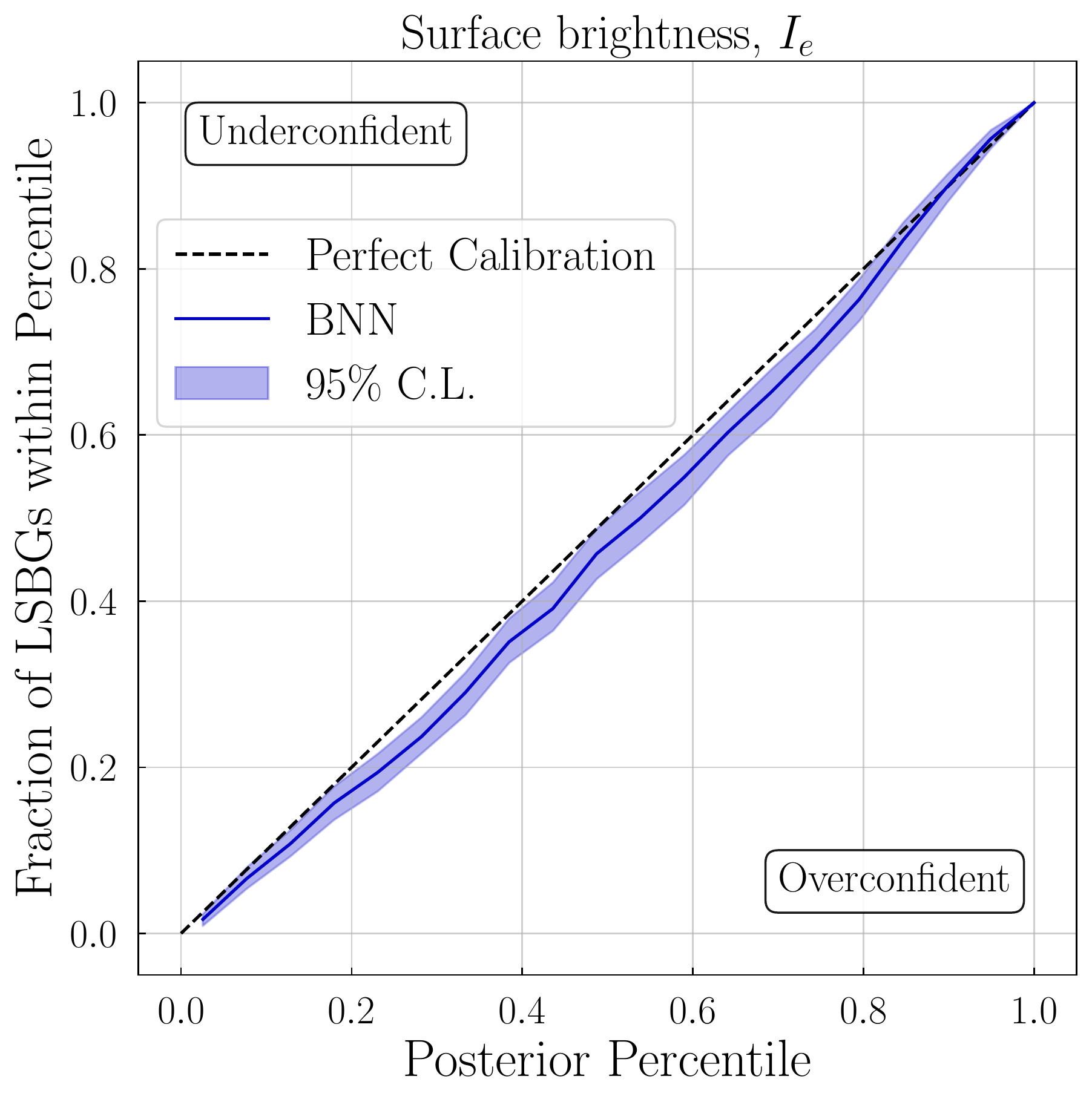}}
\hspace*{0.15cm}
\subfigure[]{\includegraphics[width=0.37\textwidth]{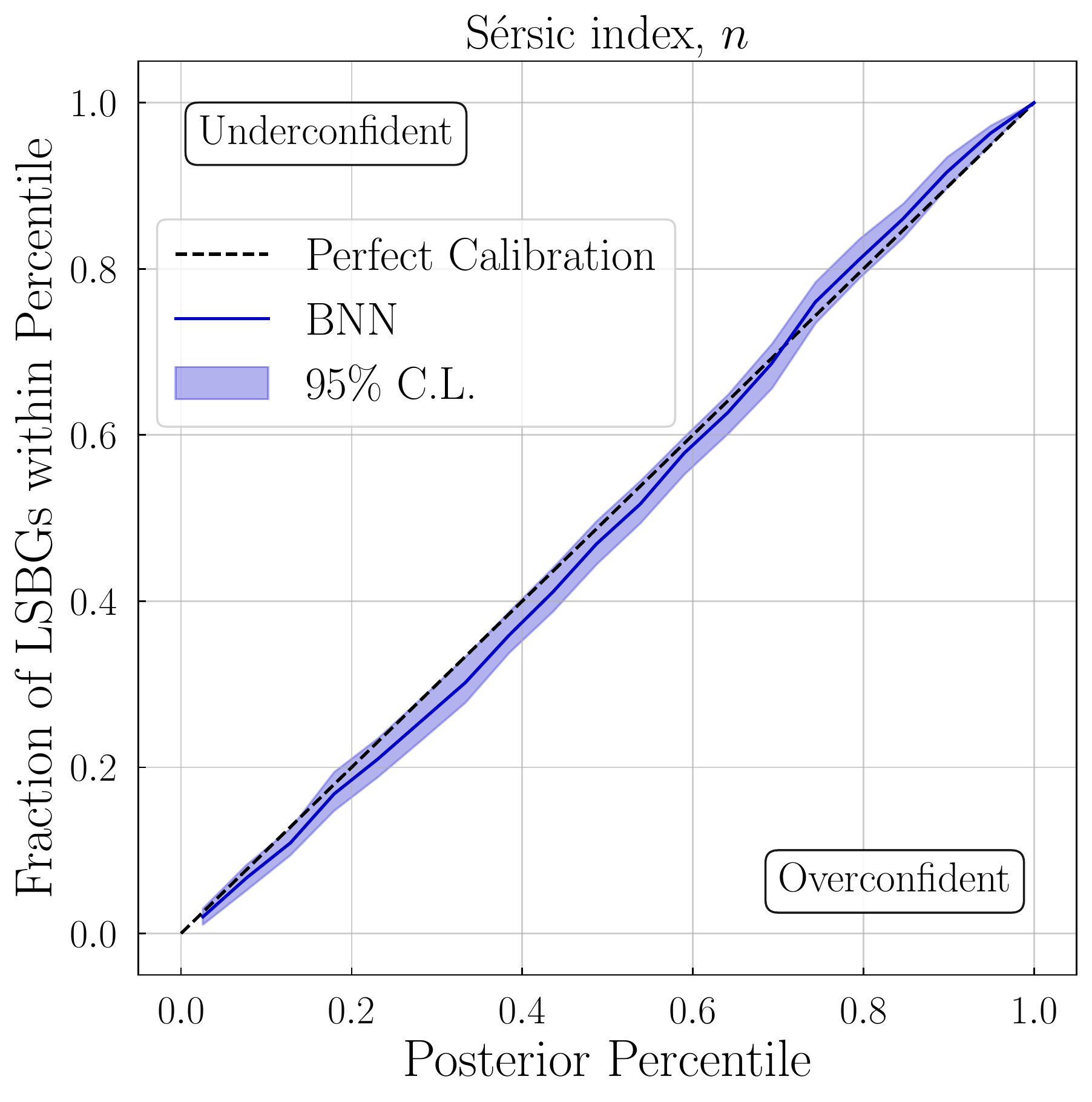}}
\vspace{-0.8cm}
\subfigure[]{\includegraphics[width=0.37\textwidth]{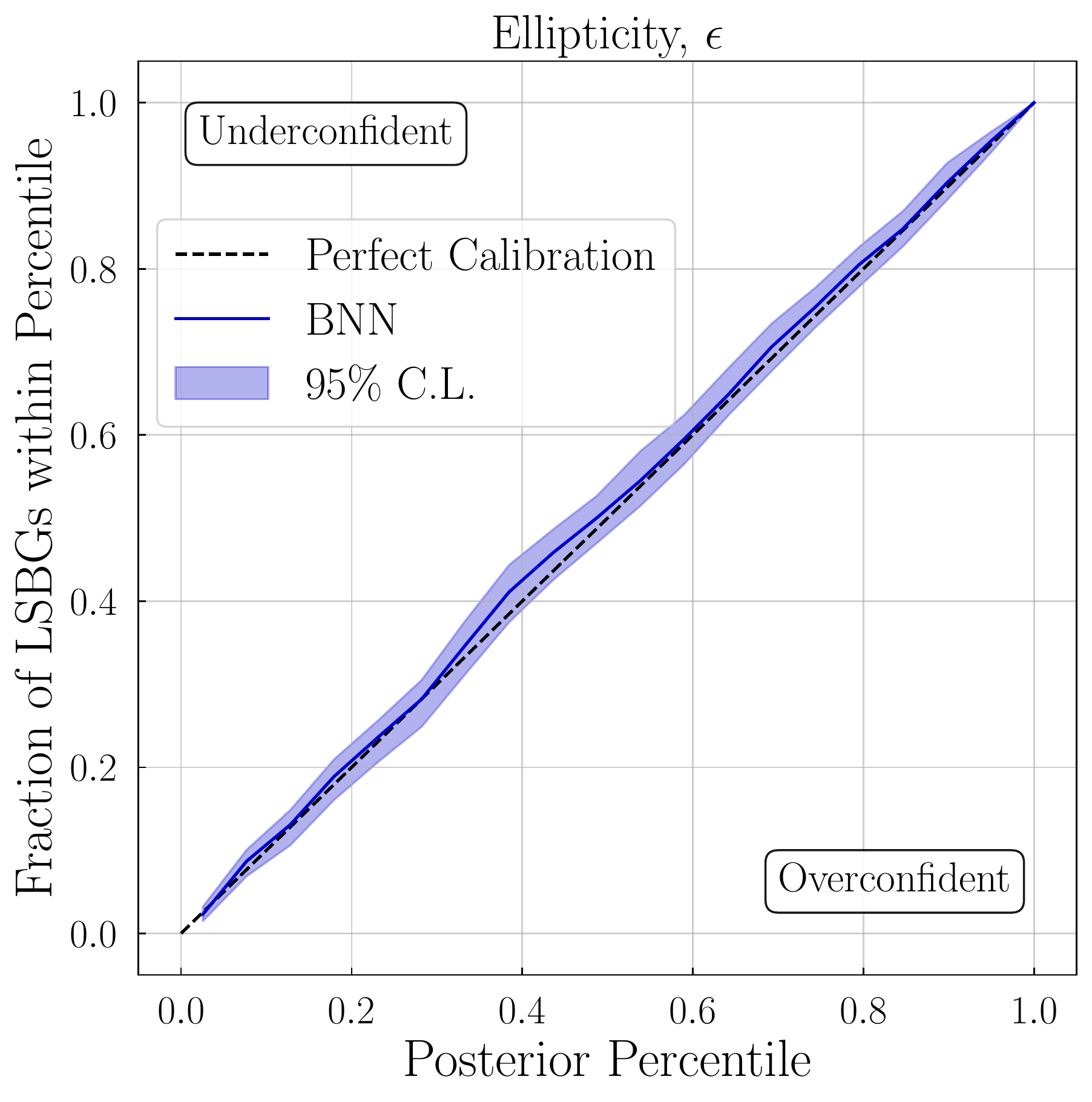}}
\hspace*{0.15cm}
\subfigure[]{\includegraphics[width=0.37\textwidth]{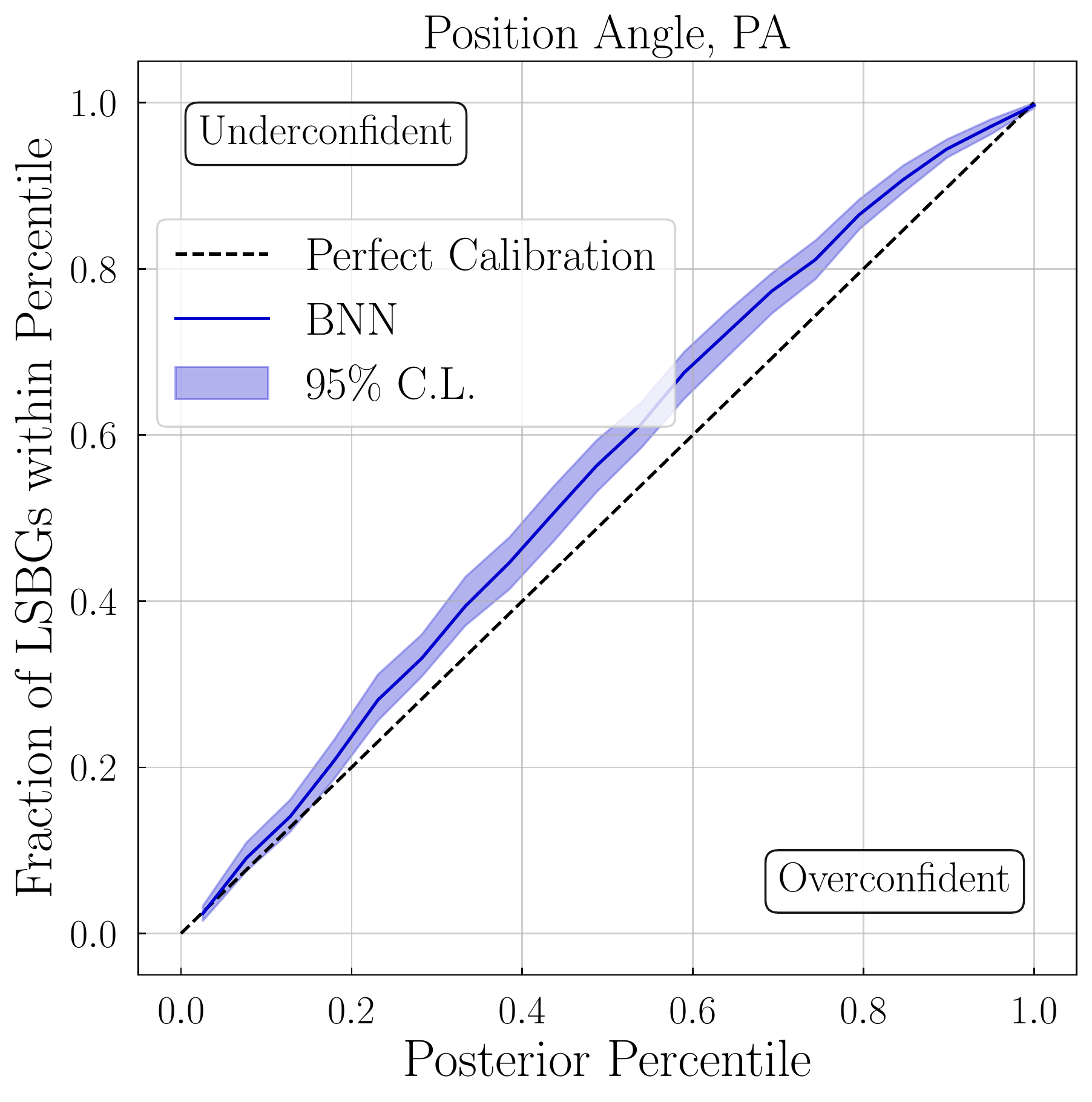}}
\vspace{0.5cm}
\caption{Calibration curves for the BNN posteriors of the four parameters of the Sérsic model, not presented in the main text. Expect for the position angle (PA) parameter, which the BNN seems to give slightly underconfident predictions, the calibration curves of the other parameters indicate perfect calibration, within the statistical uncertainty.}
\label{Calibration_Plots}
\end{center}
\vskip -0.2in
\end{figure}

\begin{figure}[ht]
\vskip 0.1in
\begin{center}
\subfigure[]{\includegraphics[width=0.37\textwidth]{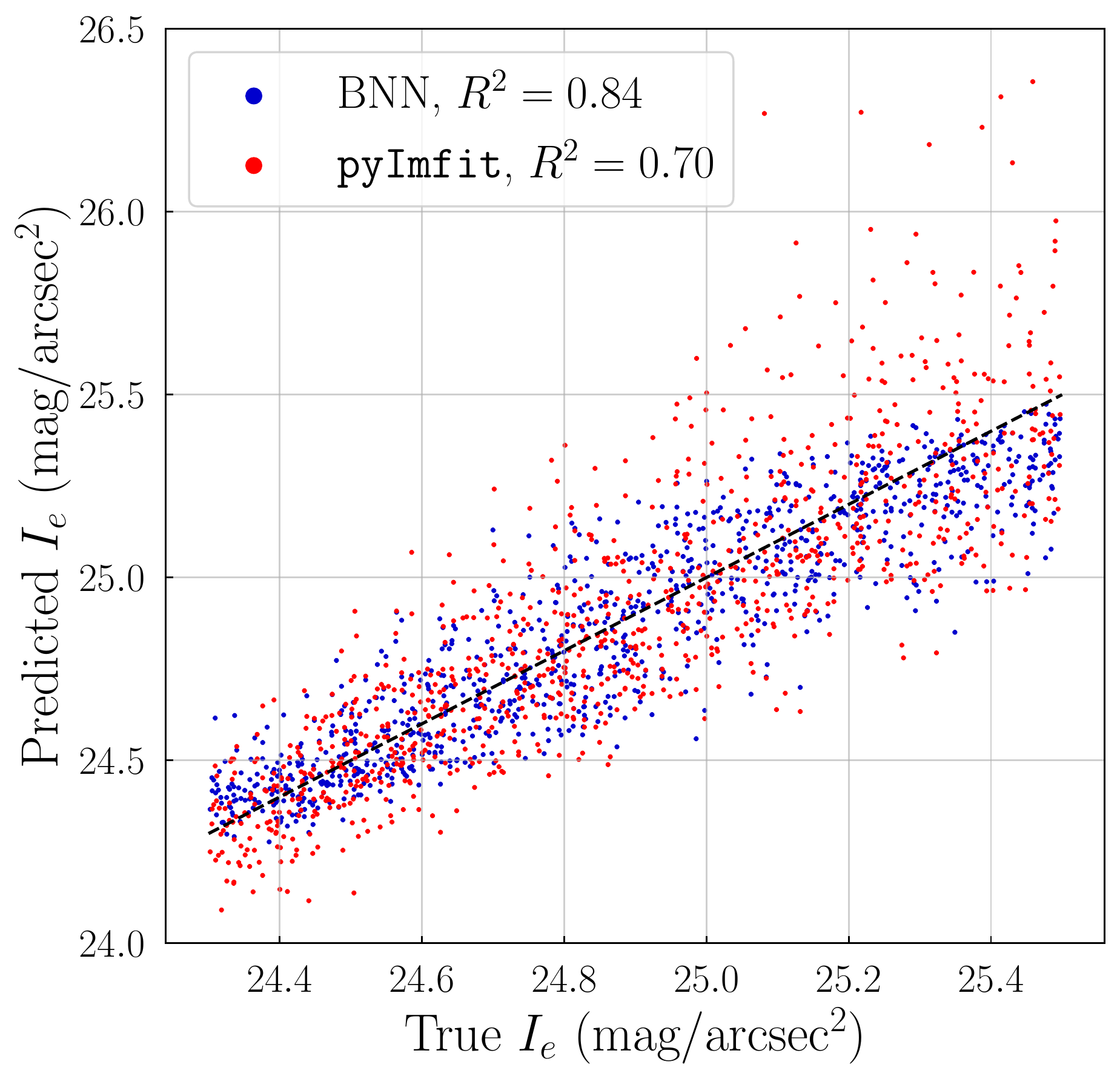}}
\hspace*{0.15cm}
\subfigure[]{\includegraphics[width=0.37\textwidth]{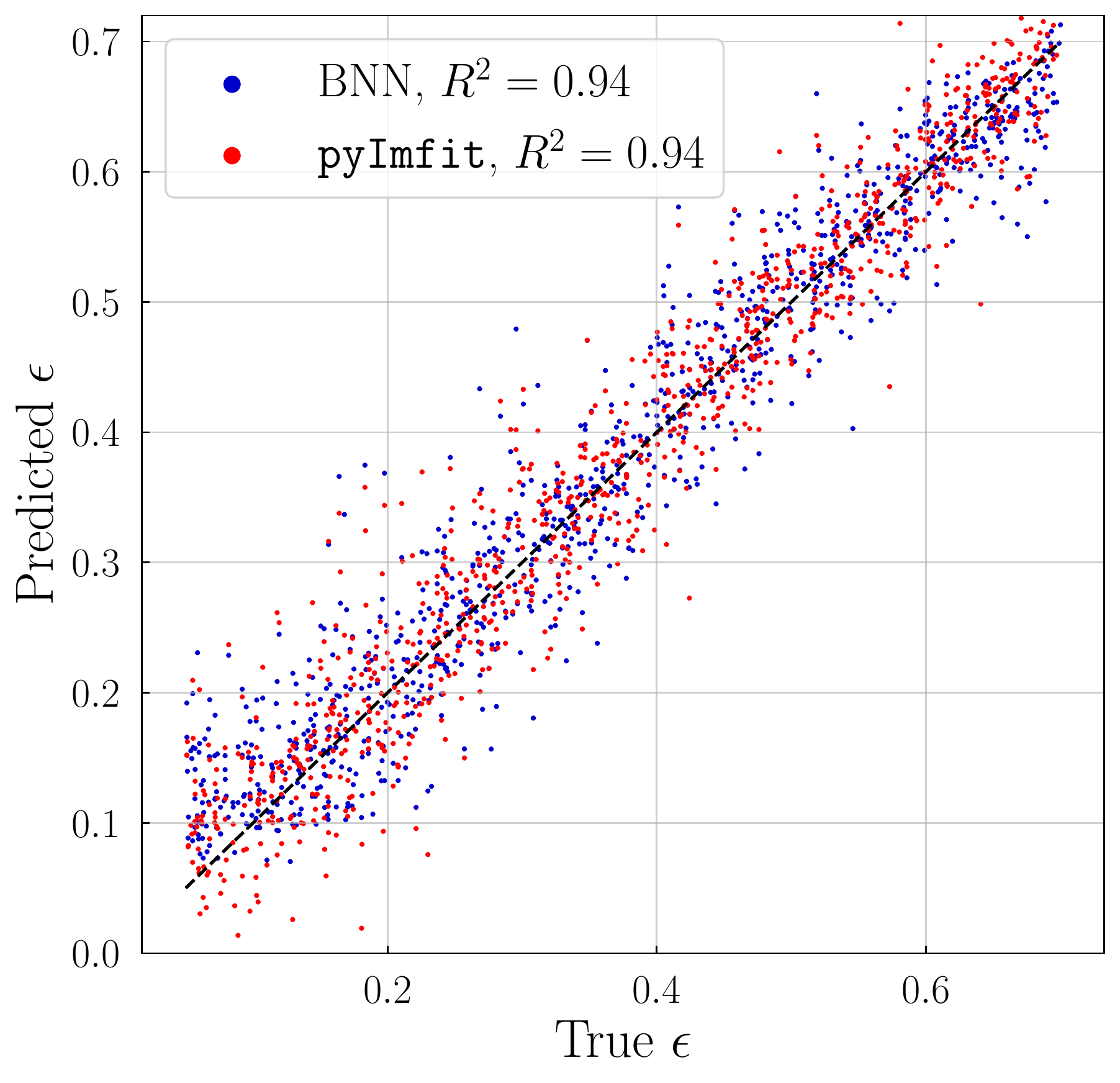}}
\vspace{-0.8cm}
\subfigure[]{\includegraphics[width=0.37\textwidth]{Appendix_Plots/BNN_pyImfit_ell.pdf}}
\hspace*{0.15cm}
\subfigure[]{\includegraphics[width=0.37\textwidth]{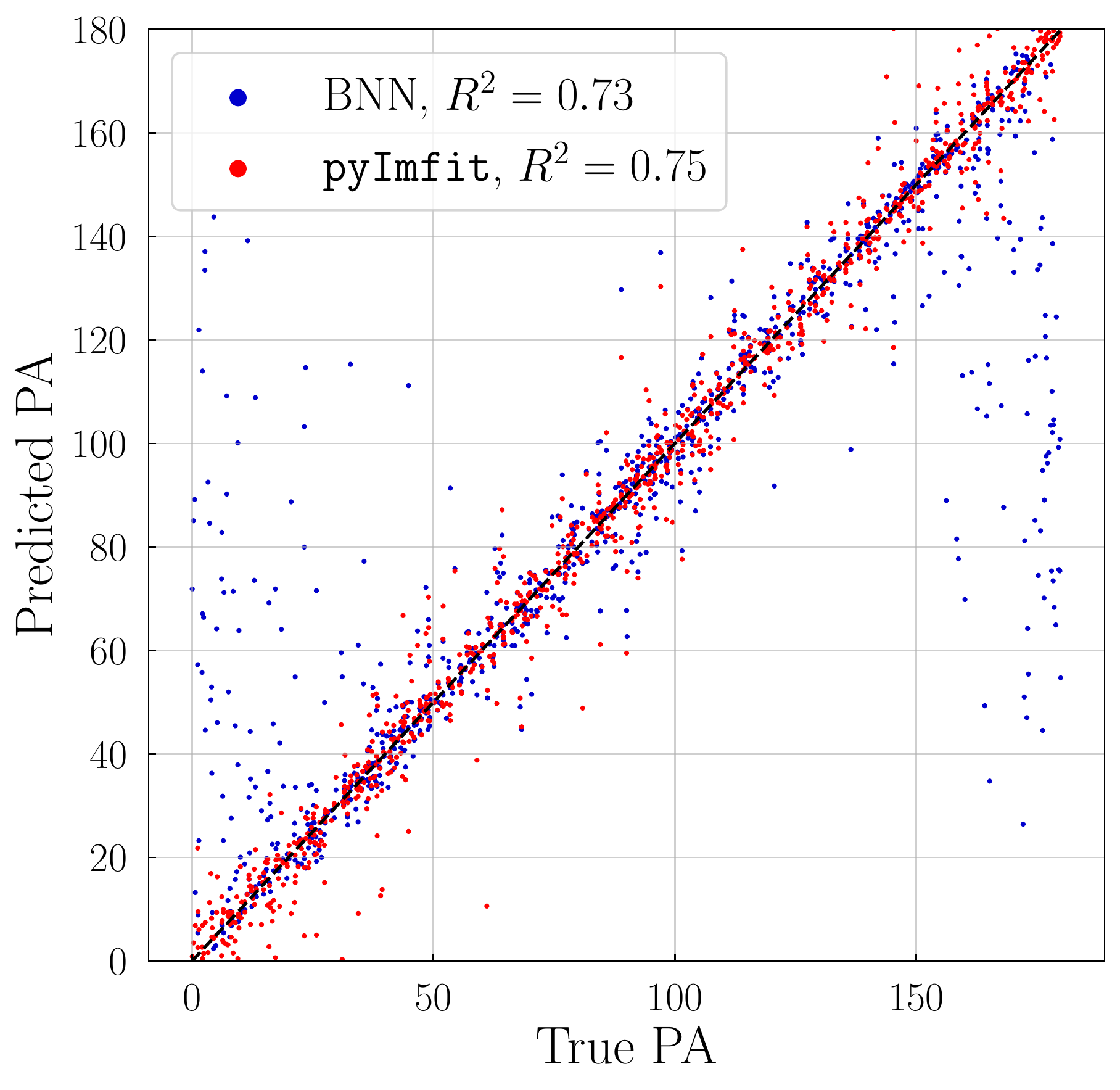}}
\vspace{0.5cm}
\caption{Comparison of the parameter predictions (point estimates) from the BNN model (blue dots) and \texttt{pyImfit} (red dots) versus the ground truth values, for the Sérsic model parameters not presented in the main text.}
\label{Point_Estimates}
\end{center}
\vskip -0.2in
\end{figure}


\end{document}